\documentclass[twocolumn]{aastex63}
\usepackage[utf8]{inputenc}
\usepackage{comment}

\newcommand{\zh}{$z_\textrm{hel}$}
\newcommand{\zc}{$z_\textrm{CMB}$}

\begin{document}
\title[Effects of Supernova Redshift Uncertainties on the Determination of Cosmological Parameters]{Effects of Supernova Redshift Uncertainties on the Determination of Cosmological Parameters} 

\author[0000-0003-3780-6801]{Charles L. Steinhardt}
\affiliation{Cosmic Dawn Center (DAWN)}
\affiliation{Niels Bohr Institute, University of Copenhagen, Lyngbyvej 2, K\o benhavn \O~2100, Denmark}

\author[0000-0002-5460-6126]{Albert Sneppen}
\affiliation{Niels Bohr Institute, University of Copenhagen, Lyngbyvej 2, K\o benhavn \O~2100, Denmark}

\author[0000-0002-1845-9325]{Bidisha Sen}
\affiliation{Cosmic Dawn Center (DAWN)}
\affiliation{Niels Bohr Institute, University of Copenhagen, Lyngbyvej 2, K\o benhavn \O~2100, Denmark}

\begin{abstract}
Redshifts used in current cosmological supernova samples are measured using two primary techniques, one based on well-measured host galaxy spectral lines and the other based on supernova-dominated spectra.  Here, we construct an updated Pantheon catalog with revised redshifts, redshift sources and estimated uncertainties for the entire sample to investigate whether these two techniques yield consistent results. The best-fit cosmological parameters using these two measurement techniques disagree, with a supernova-only sample producing $\Omega_m$ 3.2$\sigma$ higher and $H_0$ 2.5$\sigma$ lower than a hostz-only sample, and we explore several possible sources of bias which could result from using the lower-precision supernova-dominated redshifts.  In a pilot study, we show that using a host redshift-only subsample will generically produce lower $\Omega_m$ and matter density $\Omega_m h^2$ and slightly higher $H_0$ than previous analysis which, for the Pantheon dataset, could result in supernova and CMB measurements agreeing on $\Omega_m h^2$ despite tension in $H_0$.  To obtain rigorous results, though, the Pantheon catalog should be improved by obtaining host spectra for supernova that have faded and future surveys should be designed to use host galaxy redshifts rather than lower-precision methods.
\end{abstract}





\section{Introduction}
The use of Type Ia supernovae as distance indicators for constraining cosmological parameters has been essential in the development of the standard, $\Lambda$CDM cosmological paradigm \citep{Riess1998,Perlmutter1999}.  Supernovae now comprise one of a suite of tests for the expansion history of the Universe and measurements of cosmological parameters at various stages of its evolution.  Other tests include Cosmic Microwave Background (CMB) measurements \citep{Planck2018}, Baryon Acoustic Oscllation (BAO) meaurements \citep{BAOreview}, and lensed quasars \citep{H0LICOW}, as well as measurements of nearby variable objects used to calibrate some of these techniques \citep{Cepheidreview}.  

It has recently been shown that there is tension between early-Universe and late-Universe measurements of the Hubble constant $H_0$ \citep{Scolnic2014,Riess2018,Riess2019,H0LICOW,Shajib2019}.  If so, this would mean that the same $\Lambda$CDM parameters cannot simultaneously fit observations at all redshifts, requiring either a modified cosmological model or modified early-Universe physics \citep{Poulin2019,CyrRacine2016}.  These measurements of $H_0$ differ at the $\sim 10$\% level, so finding this tension has only become possible with the increased precision of recent cosmological studies.

Significant attention has been given to the statistical and systematic errors in measuring the luminosity distance to Type Ia supernova.  Uncertainties in measuring their redshifts have been given less attention, likely because those errors have often been negligible by comparison. 

It was recently pointed out that a bias in redshift estimation would produce a corresponding bias in cosmological parameters \citep{Davis2019}.  Here, we conduct a pilot study to determine whether several aspects of current redshift measurement techniques produce enough of a bias to affect the cosmological parameters derived from the Pantheon dataset.  An additional result here is that at low redshift, even {\em unbiased} redshift measurements with sufficiently large uncertainties will bias the resulting best-fit cosmological parameters.

Specifically, this work focuses on two sources of potentially significant systematic effects based on an analysis of the largest current dataset, Pantheon \citep{Scolnic2018}:
\begin{enumerate}
    \item {\bf Different Measurement Techniques:} Redshifts are determined using several different techniques, ranging from narrow emission lines in high-resolution spectra of well-identified host galaxies (hostz; with redshift uncertainties $\lesssim 0.0001$) to broad, asymmetric and outflow-dominated absorption lines in supernova spectra (SNz; with redshift uncertainties as high as $\sim 0.01$).  These measurements are combined into a common sample.
    \item {\bf Cataloging Individual Redshift Uncertainties:} The Pantheon and JLA \citep{Betoule2014} compilations report all redshift uncertainties as zero\footnote{As described below, Pantheon distance modulus uncertainties are sensitive to binned redshift uncertainties.}, and regressions using these compilations treat redshifts as zero-uncertainty measurements.  Some source catalogs used to compile these datasets report individual redshift uncertainties, but others only report whether individual redshifts were measured using a host or supernova spectrum.  This means that `errors in variables' regression techniques which would include redshift uncertainties in the fit cannot be used on current datasets.  As shown here, some redshift uncertainties can be recovered by consulting original source catalogs; but many of the spectra are not public, making it impossible to determine individual redshift uncertainties for the entire catalog.
\end{enumerate}

There is a third significant issue which is not explored in this study.  Several aspects of the Pantheon luminosity distance calibration procedure are sensitive to both the (binned) redshift distribution and individual redshifts in the catalog \citep{Rest2014,Kessler2016,Scolnic2018}.  These affect not only the reported distance moduli for each object but also reported distance modulus uncertainties, which include a component derived from redshift uncertainties.

Thus, the cosmological fits for subsets of the full catalog presented here should not be taken as conclusive measurements of cosmological parameters.  Rather, they should be taken as useful diagnostics, comparing whether subsamples of the full Pantheon catalog yield identical cosmological parameters in order to identify which of these effects may be most significant.  However, these parameters will not be properly calibrated.   

\S~\ref{sec:hostzsnz} describes the construction of Pantheon subsamples in order to investigate these potential problems.  As part of this process, we use a more precise conversion between heliocentric and CMB-frame redshifts, as well as correct a small number of transcription errors, as described in Appendix \ref{app:zcmb}.  The resulting updated catalog, which we release along with this paper, is described in \S~\ref{sec:results}.  We also show that the best-fit cosmological parameters for hostz-only and SNz-only samples disagree, with a SNz-only sample producing $\Omega_m$ 3.2$\sigma$ higher and $H_0$ 2.5$\sigma$ lower than a hostz-only sample.  In \S~\ref{sec:fitting}, we evaluate possible explanations for that disagreement.  One possible surprising result is that at low redshift, even purely statistical redshift uncertainties as low as $10^{-3}$ (typical SNz uncerainties are between $10^{-3}$ and $10^{-2}$) can produce a significant bias if these errors are neglected.  As discussed in \S~\ref{sec:discussion}, our recommendation is to only use supernovae with host redshifts for precision cosmology, and that measuring hostz should be a key requirement for future studies.  We also suggest that after doing so, a comparisons between CMB and lower-redshift measurements of $\Omega_m$ and $\Omega_m h^2$ may provide interesting cosmological probes.

\section{Sources of Pantheon Redshifts}
\label{sec:hostzsnz}

Nearly all supernovae used for cosmology were originally identified from photometric observations, with spectroscopic followup used to confirm that the supernova is a Type Ia.  That same spectrum was most commonly used to determine the redshift.  In some cases, a later followup spectrum was also taken, and for some of the most local supernovae, a host galaxy redshift from NED was used instead.  The Pantheon catalog is compiled from 12 different smaller catalogs, with each group having made individual decisions about the necessity of obtaining additional spectra and the most suitable telescopes, exposure times, and resolutions for followup spectroscopy \citep{snls, Smith2012, Sako2018, CfA1, CfA2, CfA3, CfA4, Contreras2010, ps1, clashcandels, goods, scp}.

The result is that the individual redshift measurements are of highly variable precision, with reported uncertainties varying by as much as a factor of 1000 across the sample.  The most substantial difference comes from which spectral lines were available.  Depending upon the line of sight, timing, and properties of the telescope used for followup, it was possible to obtain narrow host galaxy lines for approximately two thirds of the Pantheon supernovae, either in the original spectrum or from followup spectroscopy taken as part of later surveys.  In that case, redshifts are calculated from cross-correlating spectral features in the template and spectral features in the data.  For the remainder, only a supernova-dominated spectrum is available, with a redshift derived from strong, broad absorption lines such as Si{\sc ii} 6355.  These lines come from expanding shells exhibiting outflows at as much as $\sim$ 10000-15000 km/s \citep{Foley2011,Silverman2015}.  Templates derived from nearby supernova spectra, along with the SuperNova IDentification code (SNID) fitting package \citep{Blondin2007}, are then typically used to determine redshift. 

Because of the strong outflows and broad, sometimes skewed line profiles, redshift derived from supernova lines (SNz) have larger uncertainties than those derived from host galaxy lines (hostz).  Most, but not all, of the catalogs used to compile the Pantheon dataset describe redshifts as SNz or hostz.  A well-measured hostz might have a redshift uncertainty between $10^{-5}$ and $10^{-4}$, whereas SNz uncertainties have been assumed to be as high as $\sim 0.01$ \citep{Rest2014}.  There are also in-between cases, such as spectra dominated by the supernova with one host galaxy emission line also present.  Depending on the source catalog, these objects are sometimes classified as SNz, sometimes hostz, and sometimes described as a third category. 

It has generally been assumed that uncertainties in luminosity distance dominate, so all authors have used both sets of redshifts interchangeably, assuming that any method produces a negligible uncertainty.  Indeed, for many of the objects in the Pantheon catalog, the original source catalog did not report an individual uncertainty for each redshift measurement.  Further, because many of the followup spectra are not public, it is not currently possible to uniformly reconstruct missing redshift uncertainties across the entire sample.  As result, this work instead relies on the techniques and uncertainties reported by the original source catalogs.

\subsection{Sample Definitions}
\label{subsec:sampledefinition}

In order to test whether different redshift measurement techniques are truly interchangeable, here the Pantheon catalog is partitioned by measurement technique.  These are then fit independently in order to determine whether they yield mutually consistent cosmological parameters (\S~\ref{sec:results}).  

Similarly, because redshift uncertainties are not reported in Pantheon, currently regression techniques which incorporate errors in the independent variable cannot be used.  A subsample is constructed with well-documented individual redshift uncertainties in order to test whether these `errors in variables' regressions produce cosmological parameters consistent with those assuming redshift uncertainties are negligible.

Specifically, the subsamples used in this work are defined as follows:
\begin{itemize}
    \item {{\bf Hostz Sample}: Most redshifts are listed in source as having been determined either from host galaxy lines (hostz) or supernova lines (SNz).  The hostz sample consists of only redshifts labeled as hostz in the original source catalogs.  This includes 702 of the 1048 Pantheon objects.}
    
    \item {{\bf Not-Hostz Sample}:The remaining 346 objects are instead placed in a not-hostz sample.  Nearly all of these redshifts are labeled as SNz.  In some cases, source catalogs list a redshift as having been determined from a combination of host and supernova lines, typically without further elaboration.  Since it is unclear whether such measurements have uncertainties more typical of a hostz or a SNz, these objects are not placed in the hostz sample.  In other cases, source catalogs have used only hostz and SNz designations, and the labeling here follows the original authors.  Although these more complex objects are relatively rare, because of their inclusion this sample is labeled not-hostz rather than SNz.}
    
    \item {{\bf Known-Error Sample}: For many objects, individual redshift uncertainties are available either from the original source catalogs or other archival data.  This is most common when an independent spectrum was taken at a different time.  Thus, most of these objects lie either at very low redshift and are in NED or were part later Sloan Digital Sky Survey spectroscopic campaigns.  The high-quality known-error sample consists of only hostz measurements with well-quantified redshift uncertainties $\leq 10^{-3}$.  This includes 584 of the 1048 Pantheon objects, and is the most reliable high-quality sample constructed here.  It is also the only one suitable for regression techniques which properly consider redshift uncertainties.
    
    The remaining 464 are instead placed in an unknown-error sample in the attached catalog.  However, the unknown-error sample contains a mixture of different measurement techniques, as well as both low-precision redshifts and high-precision but insufficiently documented redshifts.  Because the unknown-error sample contains a somewhat arbitrary combination of high-quality and low-quality data, it cannot be used for meaningful analysis of either measurement quality.}
\end{itemize} 

Applying these definitions to the Pantheon catalog consists of making individual determinations about how to handle each of the constituent data sources.  A full description of the treatment of each catalog, as well as additional corrections described in \S~\ref{sec:zhtozc}, is given in Appendix \ref{app:zcmb}. 

Although the goal here is to evaluate the effects of different redshift measurement  techniques on inferred cosmological parameters, the, e.g., hostz and not-hostz samples also differ in other ways.  Hostz typically lie towards lower redshift and in galaxies where the supernova and host can more easily be resolved separately.  These differences are discussed in more detail in \S~\ref{subsec:samplediffs}

\subsection{Conversion between Heliocentric and CMB Frame Redshifts in the Pantheon Catalog}
\label{sec:zhtozc}

The Pantheon conversion between heliocentric \zh~and \zc, the redshift in the CMB frame which is used for fitting cosmology, uses a pair of low-redshift approximations.  As part of the updated catalog presented here, these have been updated to use a full relativistic treatment of velocities and corresponding redshifts, described in more detail in Appendix~\ref{app:zcmb}. 

Because the velocity of the CMB dipole is approximately 370 km/s \citep{Planck2018}, these corrections alter most of the \zc~measurements in the Pantheon catalog by $\sim 10^{-3} z$, which is on average $\sim 3 \times 10^{-4}$.  Fits of the Pantheon catalog using the original and adjusted redshifts can be compared to determine whether these redshift conversion approximations biased cosmological fits.

Using a flat $\Lambda$CDM model as an example, the best-fit $H_0$ is 0.17 km/s/Mpc higher and best-fit $\Omega_m$ 0.007 lower using corrected redshifts.  This is of similar magnitude (although with the opposite sign, likely because of the predominance of observations aligned more closely with the CMB dipole) to the effects of adding a random scatter of $\sim 10^{-3} z$ to each redshift, as discussed in more detail in \S~\ref{sec:fitting}.  These differences are smaller than fit uncertainties, and for $H_0$ nearly 10 times smaller than the dominant systematic error, which comes from calibration of the distance ladder.

Changing redshifts should also be accompanied by a corresponding shift in the \emph{K} correction \citep{Humason1956,Oke1968}.  Since distance modulus increases towards higher redshift, this effect is opposite that of errors in $z$.  At low redshift, where errors in $z$ are most significant, the effect is negligible but it can become more substantial towards $z \sim 1$.  Thus, following proper recalibration, the difference between best-fit cosmological parameters using the Pantheon redshifts and adjusted redshifts will be slightly smaller than that given above.

Regardless, this effect alone would not have induced a significant bias in the resulting cosmological fits.  However, the improved conversion increases redshift accuracy and should provide an improvement in fit quality for supernovae with very well-measured redshifts, as in those cases this redshift conversion error exceeded measurement uncertainty.  Therefore, for the remainder of this work all tests will be performed using these updated redshifts.

These redshift corrections are included in the updated Pantheon catalog (Table \ref{tab:catalog}) released along with this paper.  In addition, every object is labeled with its membership in the subsamples described in this section.  

\section{The Effects of Redshift Measurement Choices on Best-fit Cosmology}
\label{sec:results}

The effects of using different redshift measurement techniques on the Pantheon sample can be evaluated by comparing cosmological fits produced on samples restricted a single technique.  The exact effects will depend upon the particular cosmological model.  Each fit in this work uses the \texttt{scipy} orthogonal distance regression sub-package, with specific cosmological models taken from \texttt{astropy.cosmology} \citep{astropy:2013,astropy:2018}

Here, we present the best-fit parameters for the three most commonly-considered standard cosmological models (Table \ref{table:bestfit}).  Flat $\Lambda$CDM has two free parameters, $H_0$ and $\Omega_m$ and assumes that $\Omega_m + \Omega_\Lambda = 1$ and dark energy has equation of state $w = -1$.  wCDM is a flat cosmology with three free parameters, $H_0$ and $\Omega_m$ and $w$ for dark energy, and assumes that $\Omega_m + \Omega_\Lambda = 1$.  Finally, oCDM also has three free parameters, $H_0$, $\Omega_m$ and $\Omega_\Lambda$, and assumes that dark energy has $w = -1$ but does not require the Universe to be flat. 

The actual value of $H_0$ cannot be determined from supernovae alone and instead depends upon low-redshift luminosity calibration from more local rungs of the cosmic distance ladder.  Therefore fits are instead described using $\Delta H_0$, the difference between the best-fit $H_0$ for each sample and $H_0$ using the original Pantheon catalog, which is $74.03 \pm 1.43$ km/s/Mpc with the current calibration \citep{Riess2019}.  If luminosity calibration were to change substantially, the best-fit $H_0$ would also change substantially, but $\Delta H_0$ would not.  Further, the uncertainty in $H_0$ is dominated by the systematic error in luminosity calibration from the calibration of the distance ladder, but because $\Delta H_0$ does not depend upon that calibration, it can be determined far more precisely than $H_0$.

\begin{table}[ht]
\renewcommand{\arraystretch}{1.0} 
\centering
\begin{tabular*}{0.5\textwidth}{c @{\extracolsep{\fill}} ccc}
\multicolumn{4}{c}{\bf Flat $\Lambda$CDM\footnote{$\Omega_m h^2$ is calculated using the same Type Ia luminosity calibration as in \citet{Riess2019}.  The uncertainty is larger than would be implied by uncertainties in $\Delta H_0$ and $\Omega_m$ because it is dominated by the larger systematic errors in luminosity calibration.}}
\\ \hline \hline
Sample & $\Delta H_0$ & $\Omega_m$ & $\Omega_m h^2$  \\ \hline 
Hostz & $0.45 \pm 0.25$ & $0.270 \pm 0.014$ & $0.153 \pm 0.009$ \\ 
Not-Hostz & $-0.96 \pm 0.50$ & $0.374 \pm 0.029$ & $0.200 \pm 0.017$  \\ 

\hline
Known-Error & $0.49 \pm 0.29$ & $0.262 \pm 0.017$ & $0.146 \pm 0.010$ \\  
KE 2-D fit & $0.61 \pm 0.30$ & $0.257 \pm 0.017$ & $0.144 \pm 0.009$ \\ 
\hline

\multicolumn{4}{c}{\bf wCDM}
\\ \hline \hline
Sample & $\Delta H_0$ & $\Omega_m$ & $w$  \\ \hline 
Hostz & 0.55 $\pm$ 0.35 & 0.29 $\pm$ 0.05 & -1.05 $\pm$ 0.15 \\
Not-Hostz\footnote{The fit does not converge, but rather is optimized over the allowed domain at $\Omega_m = 0,~\Omega_\Lambda =1.$} & -2.36 $\pm$ 1.00 & 0 & -0.44 $\pm$ 0.36 \\
\hline
Known-Error & 0.41 $\pm$ 0.51 & 0.25 $\pm$ 0.08 & -0.96 $\pm$ 0.19 \\
KE 2-D fit & 0.55 $\pm$ 0.52 & 0.25 $\pm$ 0.08 & -0.97 $\pm$ 0.19 \\ 
\hline
\multicolumn{4}{c}{\bf oCDM}
\\ \hline \hline
Sample & $\Delta H_0$ & $\Omega_m$ & $\Omega_\Lambda$  \\ \hline 
Hostz & 0.23 $\pm$ 0.32 & 0.30 $\pm$ 0.05 & 0.75 $\pm$ 0.08 \\
Not-Hostz & -2.38 $\pm$ 1.16 & 0.21 $\pm$ 0.13 & 0.27 $\pm$ 0.28 \\

\hline
Known-Error & 0.41 $\pm$ 0.45 & 0.25 $\pm$ 0.08 & 0.71 $\pm$ 0.12 \\ 
KE 2-D fit & 0.56 $\pm$ 0.45 & 0.25 $\pm$ 0.08 & 0.72 $\pm$ 0.12 \\ 
\hline 
\end{tabular*}
    \caption{Best-fit cosmological parameters for three different models and the samples described in \S~\ref{sec:hostzsnz}. Cosmological models are described in \S~\ref{sec:results}.  All fits assume redshift uncertainties are negligible with the exception of the "2-D" fit for the Known-Error (KE) sample, which uses reported uncertainties.  The hostz and not-hostz samples yield inconsistent cosmological parameters, with the not-hostz $\Omega_m$ 3.2$\sigma$ higher and $H_0$ 2.5$\sigma$ lower than the hostz sample.}
    \label{table:bestfit}
\end{table}

Like the original Pantheon sample, the best-fit wCDM model for the high-precision hostz and known-error samples is consistent with $w = -1$ dark energy and the best-fit oCDM model is consistent with a flat cosmology, $\Omega_m + \Omega_\Lambda = 1$.  Therefore, the remainder of the discussion will focus on the best-fit flat $\Lambda$CDM cosmology.

\begin{figure}[htp] 
    \centering
    \includegraphics[width=0.5\textwidth]{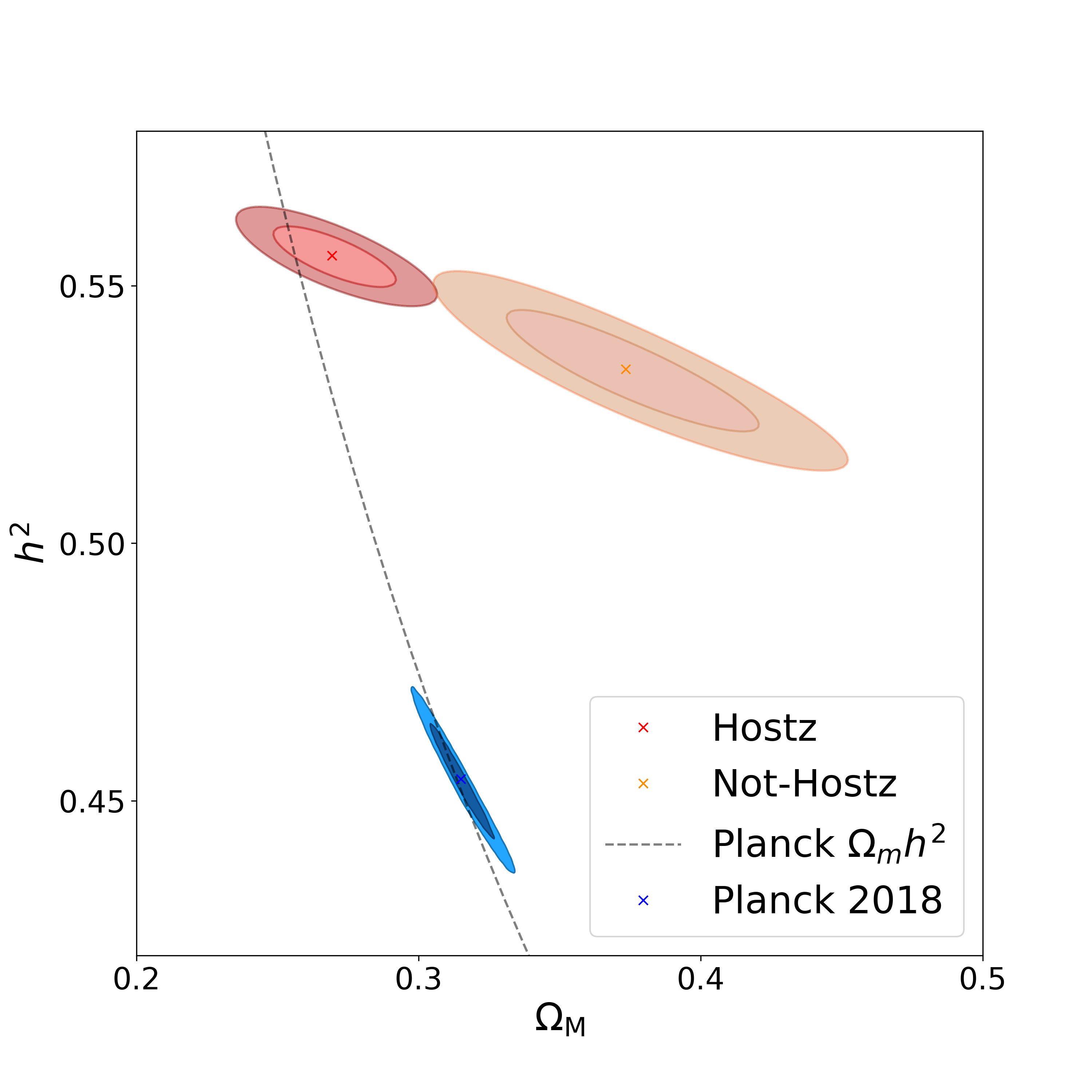}
    \caption{68.3 $\%$ and 95.4 \% confidence regions of the ($h^2$,$\Omega_m$) plane from hostz SNe, not-hostz SNe and Planck 2018 for a flat $\Lambda$CDM cosmology. The hostz and not-hostz sample differ by slightly over 3$\sigma$ in inferred cosmological parameters.  Cosmological parameters inferred from the hostz sample are consistent with the matter density $\Omega_m h^2$ measured at early times.}
    \label{fig:parameters}
\end{figure}

The two hostz samples, hostz and known-error, have considerable overlap with each other, and therefore agree to higher precision than the fit uncertainty.  If hostz and SNz measurements are interchangeable, they should also have agreed with the not-hostz sample.  However, there is instead disagreement between the hostz and not-hostz fits.  The not-hostz sample finds $\Delta H_0$ 2.5$\sigma$ higher and $\Omega_m$ 3.2$\sigma$ higher than hostz.  Fitting cosmological parameters to Type Ia supernovae with redshifts from host galaxy lines and those with redshifts from a supernova-dominated spectrum produces two different results (Fig. \ref{fig:parameters}).

Because these discrepancies are only at the $2.5 -- 3.2 \sigma$ level, it would be unclear whether this alone is prove that hostz and SNz are not interchangeable.  A key point of this work is that the division into hostz and SNz is not arbitrary.  Rather, it is known that systematic errors in redshift could create such an effect \citep{Davis2019} and a comparison of hostz and SNz for a sample of SDSS objects indicates a systematic difference of $2.2\times 10^{-3}$.  In this work, it is also shown that statistical uncertainties alone can produce a similar effect.

In \S~\ref{sec:fitting}, these effects are explored further, showing that the likely result of greater errors in SNz would be a larger $\Omega_m$ and smaller $H_0$ than the hostz sample.  Because host redshifts are used wherever possible, the SNz sample has different selection and includes host galaxies with different properties than the hostz sample.  The tests done in this work cannot isolate these variables, a study which will likely require additional observations.  However, it should now be clear that these two techniques should not be combined into a common catalog; at least one of the two currently produces a systematically incorrect result.

\begin{figure}[ht] 
    \centering
    \includegraphics[width=0.5\textwidth]{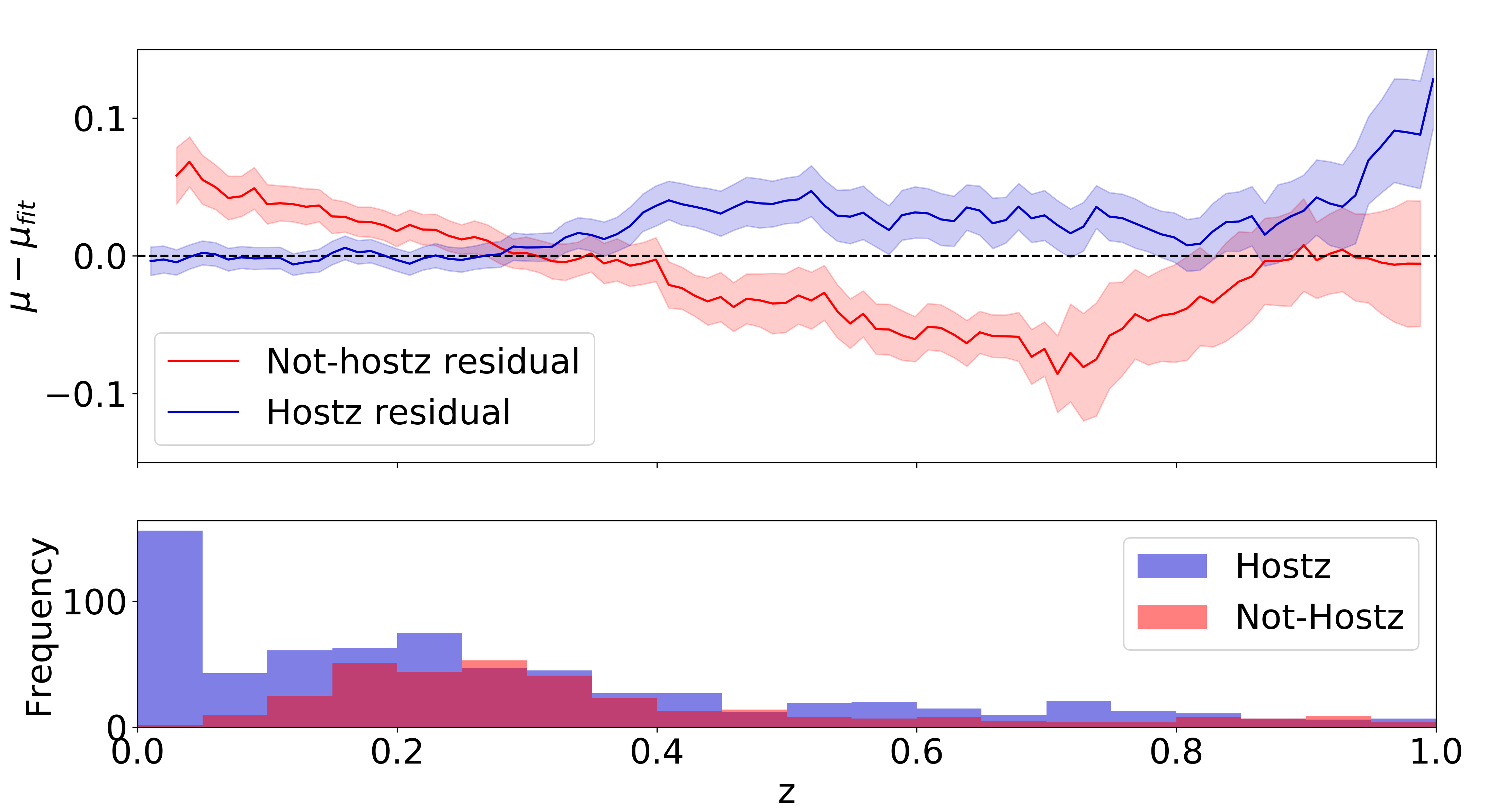}
    \caption{Boxcar-smoothed residuals with bin size of 0.2 for not-hostz sample (red) and hostz sample (blue) compared with the best-fit not-hostz cosmology.  Estimated statistical uncertainties on the smoothed residuals are calculated using resampling and shown as lighter shading.  The not-hostz residuals exhibit the same characteristic shape as the biases illustrated in Figs. \ref{fig:residualfits} and \ref{fig:residualfitssys}, although a statistically rigorous comparison is not possible due to unknown redshift uncertainties.  Further, additional selection differences between the samples discussed in \S~\ref{subsec:samplediffs} will also contribute to the shape of the residuals.  Redshift distributions of the hostz and not-hostz samples are shown at bottom.  Lower-redshift hosts are far more likely to have been resolved independently or to have had existing host spectra.}
    \label{fig:residualcomp}
\end{figure}

Of the two, it appears that the hostz sample is better described by $\Lambda$CDM models than the SNz sample.  Although individual redshift uncertainties are not available, and thus a proper $\chi^2$ comparison is not meaningful, there are several other indicators which hint at this conclusions.  The SNz fit does not converge at all for the wCDM model; the regression terminates at $\Omega_m = 0,~\Omega_\Lambda = 1$, with a negative $\Omega_m$ disallowed as unphysical.  Further, a comparison of the residuals to respective best-fit cosmologies hints at the same conclusion (Fig. \ref{fig:residualcomp}): the hostz and not-hostz residuals to the best-fit combined cosmology diverge, and the combined cosmology is not a good description of either sample.  The not-hostz sample residuals exhibit a shape characteristic of the biases illustrated in Figs. \ref{fig:residualfits} and \ref{fig:residualfitssys}, with the latter likely the dominant contribution.\footnote{A statistically rigorous comparison is not possible due to the unknown redshift uncertainties.}, described in more detail in the following section.  Additional selection differences between the samples discussed in \S~\ref{subsec:samplediffs} will also contribute to the shape of the residuals.

The results are more promising from testing the assumption that well-measured redshifts have negligible uncertainties.  Because reliable uncertainties are only available for the known-error sample, only the "KE 2-D" fits in Table \ref{table:bestfit} use reported redshift uncertainties.  These best-fit parameters are consistent with those using a "1-D" technique which ignores errors in the independent variable.  The most significant impact of this difference in fitting technique is on $\Delta H_0$, since it is most sensitive to the lowest redshift objects where small uncertainties are significant.  However, even at low redshift, the $< 10^{-3}$ uncertainties for the known-error sample indeed appear to be negligible.  

Finally, it should be stressed that restriction from the full catalog to these subsamples means that completeness differs from the full Pantheon sample. This presents a problem, because the Pantheon distance modulus uncertainties include components for effects such as redshift scatter and simulated completeness (cf. \citealt{Kessler2016} for the Pantheon sample and \citealt{Kessler2009} for the SNANA package which was used), as well several sources of systematic uncertainty.  During this process, supernova are binned by redshift, with redshifts of all different acquisition methods treated as interchangeable.  Running these same techniques on only a subsample of the Pantheon catalog would therefore not merely remove some points, but also alter the ones which remain.  As a result, subsets of the Pantheon sample, such as the ones given here, are not properly calibrated for determining cosmological parameters. The known-error sample, with its restriction to well-documented hostz measurements, also treats different source surveys in different ways, adding additional possible biases which are difficult to evaluate.  

All of these must be addressed before best-fit cosmological parameters can be considered reliable.  The fits in Table \ref{table:bestfit} demonstrate that hostz and SNz yield different cosmologies, and hostz are indeed more precise than SNz.  However, the hostz subsample included in the catalog here, while likely better than the combined sample, is not suitable for meaningful cosmological parameter determination without additional calibration.  For the same reason, it is difficult to quantify the effects of systematic uncertainties, since in the Pantheon sample some of these are included as a component of the total uncertainty listed.  For $H_0$, the dominant systematic uncertainty will almost certainly still come from calibration of the cosmic distance ladder, but for other parameters, a careful separation of statistical and systematic errors would be required.  Indeed, what we show here is that the dominant systematic uncertainty in the determination of $\Omega_m$ appears to be the relative fraction and redshift distribution of hostz and SNz in the supernova sample.

\section{Explanations for the Disagreement Between Hostz and SNz Samples}
\label{sec:fitting}

There are three principal differences between the hostz and SNz samples, each of which may contribute to the disagreement between hostz-only and SNz-only cosmological fits:
\begin{itemize}
    \item SNz have significantly larger statistical uncertainties, ranging from 10-1000 times higher than hostz.
    \item SNz require modeling supernova outflows and broad, asymmetric lines.  This is a more difficult problem than fitting narrow galaxy lines, both because of the line widths and because unlike host galaxies, choosing the wrong template can produce systematic redshift errors.
    \item SNz host galaxies lie at higher redshift and have different morphology and other properties than hostz galaxies.
\end{itemize}
The mechanisms by which all three might bias the resulting cosmology are explored in more detail below. 

The approach taken in this section is first to demonstrate the mechanisms by which errors in SNz can, in principle, bias cosmological parameters for a supernova sample using very simple assumptions.  Statistical uncertainties (\S~\ref{subsec:statistical}) are shown to be important primarily for low-redshift objects, whereas systematic uncertainties (\S~\ref{subsec:systematic}) can create a bias even for a high-redshift sample.  Then, the actual redshift distribution and uncertainties of the Pantheon sample are used to determine whether they are large enough to explain the discrepancy between hostz and not-hostz results shown in \S~\ref{sec:results}.  Because most low-redshift objects in the Pantheon catalog have hostz measurements, the bias between hostz and not-hostz parameters is likely dominated by systematic, rather than statistical, uncertainties.

\subsection{Statistical Uncertainties and the Determination of Cosmological Parameters}
\label{subsec:statistical}

The bias from larger statistical uncertainty in SNz redshifts does not stem from the uncertainties themselves, but rather a fitting routine which treats redshifts as exact.  Using, e.g., orthogonal least squares regression to fit cosmological parameters for a simulated dataset with even very large redshift uncertainties will still return the input cosmology.  However, neglecting errors in the independent variable and using a one-dimensional regression technique will yield a biased result if there are errors in both variables \citep{Isobe1990,Feigelson1992,Steinhardt2018}.  

\begin{figure}[ht] 
    \centering
    \includegraphics[width=0.5\textwidth]{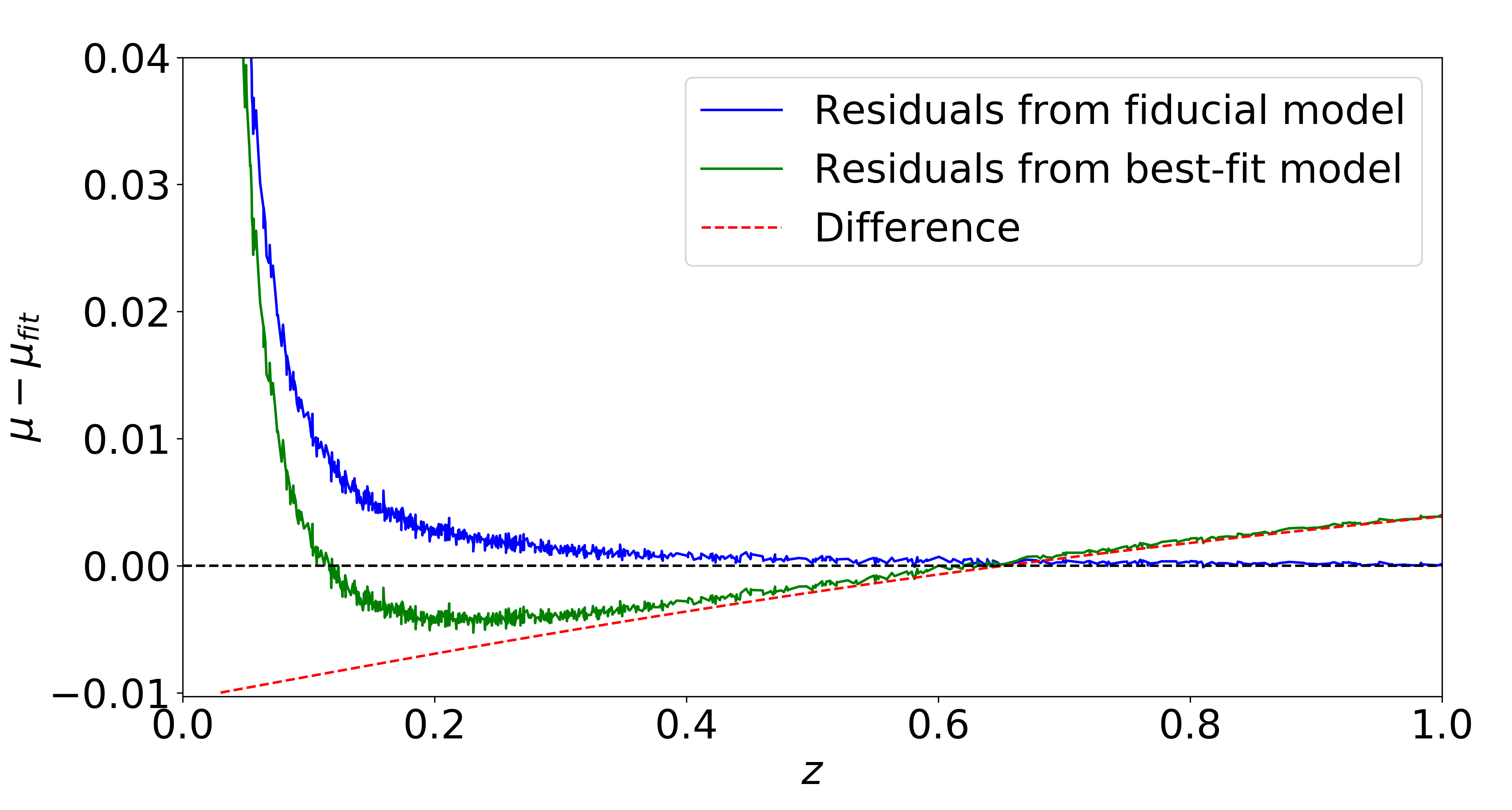}
    \caption{The effects of a statistical redshift uncertainty of $0.01$ on a sample drawn from the Pantheon $z > 0.03$ redshift distribution.  Each object is scattered over 10,000 trials, producing a positive average residual compared against the input cosmology which sharply increases towards low redshift (blue).  For relatively small changes in parameters, the difference between $\mu(z)$ for the best-fit flat $\Lambda$CDM cosmology to scattered points is approximately linear (red).  The only residual which can actually be measured is the difference between the scattered points and the best-fit cosmology (green), since the input cosmology is unknown for real data.}
    \label{fig:residualfits}
\end{figure}

For supernova cosmology, there is an additional effect because $\mu(z)$ is observed to be sub-linear.  As a result, an object scattered downward by $\Delta z$ will overestimate the correct $\mu$ at its new redshift by more than an object scattered upward by that same $\Delta z$ will underestimate it, creating an additional bias.  This bias goes up sharply towards low $z$, where $d^2\mu/dz^2$ is largest (Fig. \ref{fig:residualfits}a, blue).

These scattered points will no longer be best-fit by the cosmology from which they were drawn, but rather by one with different parameters.  For flat $\Lambda$CDM cosmologies, this flatter slope at low redshift corresponds to a lower $H_0$ and matching the high-redshift end thus requires a higher $\Omega_m$.  For small changes in parameters, the difference between the old and new $\mu(z)$ is approximately linear (Fig. \ref{fig:residualfits}, red).  The cut at $z > 0.03$ used to produce Fig. \ref{fig:residualfits} in is useful for illustration purposes, but also produces a difference in best-fit cosmology smaller than for the full Pantheon sample with a similar cut instead at $z < 0.01$.

\begin{figure}[ht] 
    \centering
    \includegraphics[width=0.48\textwidth]{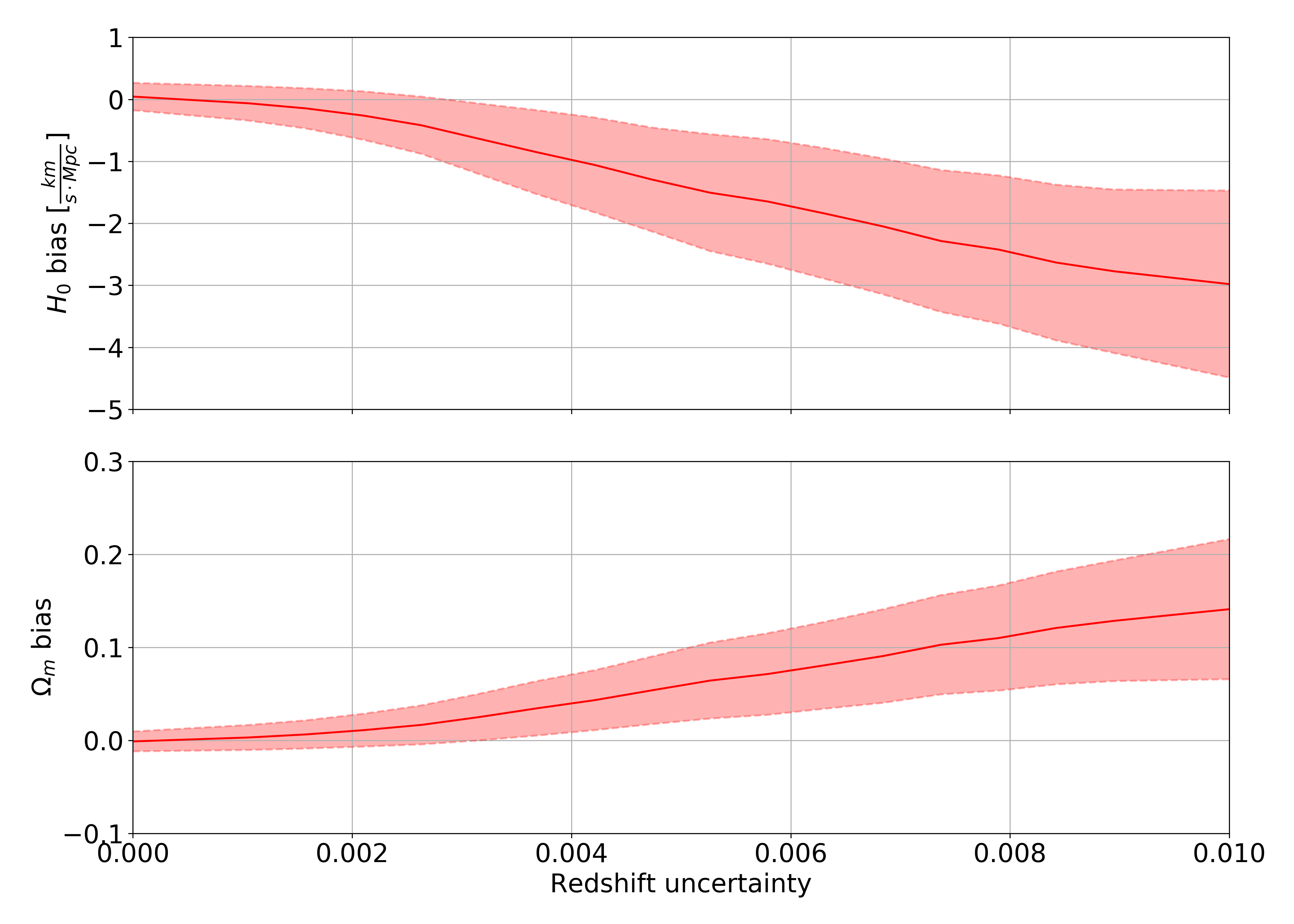}
    \caption{Difference between the best-fit values of $H_0$ and $\Omega_m$ for flat-$\Lambda$CDM and input Planck cosmology for simulated regressions ignoring uncertainties in redshift for a sample with a Pantheon-like redshift distribution.  At a redshift uncertainty of $\sim 10^{-3}$, the regression begins to produce a biased fit, underestimating $H_0$ and overestimating $\Omega_m$.  Greater redshift uncertainty leads to a larger bias as well as a larger scatter between different trials (1$\sigma$ contours shown).  Thus, the not-hostz sample will be biased towards small $H_0$ and large $\Omega_m$.  The full Pantheon sample is composed of a combination of the hostz and not-hostz samples and will be less biased than the not-hostz sample.  The exact bias is sensitive not only to the redshift distributions of the two samples (Fig. \ref{fig:residualcomp}, bottom) but also the individual redshift uncertainties.}
    \label{fig:onedfit}
\end{figure}

The effects on best-fit cosmological parameters are shown in Fig. \ref{fig:onedfit}.  As expected, an increase in statistical uncertainty leads to a decrease in the best-fit $H_0$ and increase in $\Omega_m$, biasing both with respect to the input cosmology.  Even though the actual value of $H_0$ cannot be determined directly from supernovae and depends on the calibration, it is one of the fit parameters.  Increasing the redshift uncertainty will decrease the best-fit $H_0$, so that fitting cosmological parameters using the correct distance ladder calibration will not produce the correct $H_0$, but rather one which is lower.  

If all redshift errors in the entire pantheon sample were as large as $10^{-2}$, the largest redshift uncertainty reported for any individual measurement (\S~\ref{sec:hostzsnz}), the resulting $\Omega_m$ would increase by 0.15 and $H_0$ would decrease by 3 km/s/Mpc (as seen in Fig. \ref{fig:onedfit}).  However, the bias introduced by a redshift scatter is dominated by the lowest redshift SNe as seen in Fig. \ref{fig:residualfits}. In practice, redshift uncertainties are not uniform across the Pantheon catalog, and modeling the resulting bias is more complicated. In particular, the lowest-redshift galaxies are most likely to have hostz and thus be well measured (Fig. \ref{fig:residualcomp}, bottom) and the not-hostz distribution instead peaks at $0.15 < z < 0.35$.  If redshift uncertainties only in the not-hostz sample were $10^{-2}$, with the remainder perfectly measured, the resulting $\Omega_m$ would only increase by 0.03 and $H_0$ would decrease by 0.7 km/s/Mpc.  Here the observed bias in $\Omega_m$ is significantly less than the difference between hostz and not-hostz samples seen in Table \ref{table:bestfit}, and the expected $\Delta H_0$ is similar to the observed difference in Table \ref{table:bestfit}.   

For uncertainties below $\sim 10^{-3}$, the bias is negligible.  This explains why the known-error and "KE 2-D" fits (Table \ref{table:bestfit}) were consistent.  SNz measurements, with uncertainties as high as $10^{-2}$, will be biased, but the effect should be negligible for hostz measurements.  

It should be noted that underestimating measurement uncertainties will therefore lead not just to a formally invalid result, but also to an underestimation of fit uncertainties.  This is particularly important when comparing supernova measurements of $H_0$ and $\Omega_m$ with other measurements, since the presence of tension depends on the uncertainties. For $H_0$, the fit uncertainty is still negligible compared with uncertainties arising from calibration of the distance ladder.  For $\Omega_m$, the fit uncertainty is more significant.  

\subsection{Systematic Errors at Higher Redshifts}
\label{subsec:systematic}

The lowest-redshift objects in the Pantheon catalog typically have host spectra, and generally higher-quality redshift determination.  Thus, relatively few objects at low redshift are SNz with uncertainties that could approach $0.01$.  Even these few objects can be significant, because the residual from scattering an object at $z = 0.05$ by $0.01$ is approximately ten times that from an object at $z = 0.15$, and a least-squares minimization depends upon the squares of the residuals.  Indeed, the few lowest-redshift objects drive most of the bias in Fig. \ref{fig:residualfits}.  Thus, at a minimum one should be concerned about a sample which includes even a small number of poorly-measured redshifts at low $z$.  

\begin{figure}[ht] 
    \centering
    \includegraphics[width=0.48\textwidth]{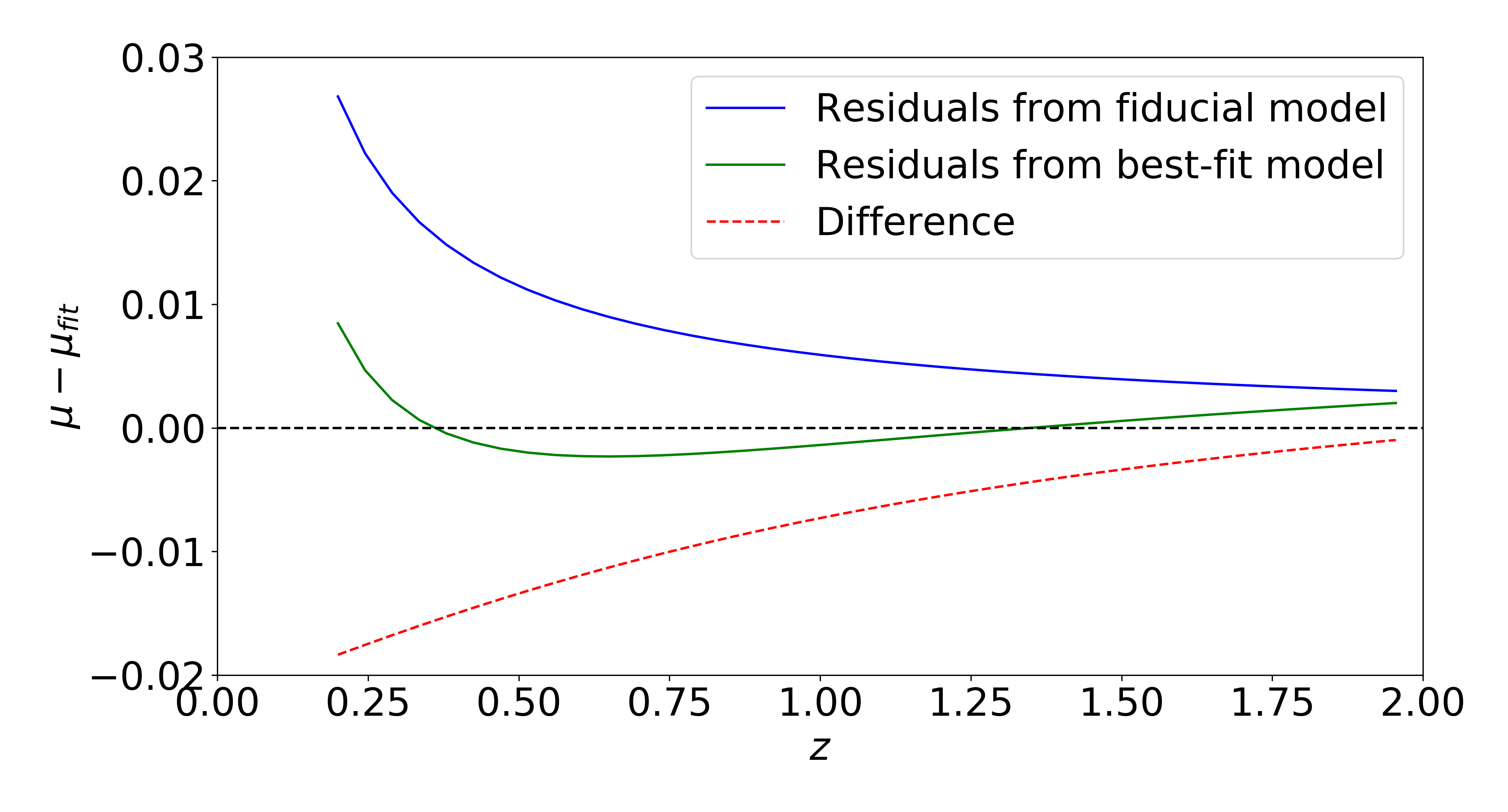}
    \caption{Corresponding effects to Fig. \ref{fig:residualfits} for an evenly-distributed sample over $0.2 < z < 2.0$ for which redshifts show a systematic error of $0.0022$ as reported for SNz in \cite{Sako2018}. This bias produces a residual compared against the input cosmology which sharply increases towards low redshift (blue).  For relatively small changes in parameters, the difference between $\mu(z)$ for the best-fit flat $\Lambda$CDM cosmology to scattered points is approximately linear (red).  The only residual which can actually be measured is the difference between the scattered points and the best-fit cosmology (green), since the input cosmology is unknown for real data.}
    \label{fig:residualfitssys}
\end{figure}

All of these uncertainties should become negligible at sufficiently high redshift.  However, a systematic error of $0.001$, ten times smaller than the reported statistical uncertainty, would have similar effects even if restricted to higher redshifts (Fig. \ref{fig:residualfitssys}; see also the extended discussion in \citealt{Davis2019}).  This is of particular concern because SNz typically rely on modeling large outflows which are $\sim 10000$ km/s \citep{Foley2011,Silverman2015} and can be skewed, and a systematic error of $\sim 300$ km/s in modeling those outflows would produce a redshift bias of $\sim 0.001$.

In order to test for such a possible bias, one can consider cosmological parameters derived only from higher-redshift objects where the statistical uncertainty is truly negligible.  Fitting the $z > 0.2$ portion of the not-hostz sample independently yields $\Delta H_0 = -0.64 \pm 0.73$, $\Omega_m = 0.360 \pm 0.36$, and $\Omega_m h^2 = 0.194 \pm 0.020$.  The $z > 0.2$ portion of the hostz sample yields $\Delta H_0 = 0.71 \pm 0.56$, $\Omega_m = 0.260 \pm 0.23$, and $\Omega_m h^2 = 0.145 \pm 0.014$.  Even at high redshift, objects without hostz measurements behave like the complete not-hostz sample and objects with hostz behave like the complete hostz sample.  Thus, even over a redshift regime where statistical uncertainties $\sim 10^{-2}$ should become negligible, the discrepancy between hostz and not-hostz fits remains.

One possible explanation is that SNz have systematic errors $\gtrsim 10^{-3}$ in addition to their far larger statistical uncertainties.  \citet{Sako2018} found that the mean offset between SNz and hostz for the SDSS sample is $2.2\times 10^{-3}$, which would be consistent with this explanation. Combining such a systematic uncertinty with estimated statistical uncertainties would produce biases in both $H_0$ and $\Omega_m$ similar to the difference between hostz and not-hostz fits in Table \ref{table:bestfit}.
However, because there are other systematic differences between the hostz and not-hostz samples, properly testing whether this is the primary cause will likely require measuring host galaxy redshifts for the remainder of the not-hostz sample, then comparing the resulting fits.

In principle, one could maximize the sample size by removing only the SNz which produce the most bias.  In practice, this is difficult because authors typically do not report individual uncertainties.  Many of the individual papers from which Pantheon is compiled just list all SNz as having an uncertainty of dz = $10^{-2}$, for example, although SNID fits vary in precision by at least an order of magnitude and individual uncertainties might be anywhere from $\lesssim 10^{-3}$ to $\sim 2\times 10^{-2}$.  Here, rather than attempt to find a small subset of SNz which can be used without introducing significant bias, we choose the more conservative approach of rejecting all SNz.  A potential danger of this approach, discussed in the following section, is that this also selects against supernovae in faint hosts, which creates an increasingly strong selection bias towards high redshift.

Finally, additional systematic effects arise because the Pantheon dataset also cuts all objects at $z < 0.01$ and rejects measurements which are 3$\sigma$ outliers from the best-fit cosmology.  Six outliers are rejected from 1054 measurements on this basis, which is consistent with purely statistical fluctuations (should occur 6\% of the time).  In a least-squares minimization, these outliers both have significant influence on the fit and on the resulting error estimates.  Both cuts lead to an additional bias when redshift uncertainties are sufficiently large, but are negligible for redshift uncertainties $\lesssim 10^{-3}$.

\begin{figure}[ht] 
    \centering
    \includegraphics[width=0.48\textwidth]{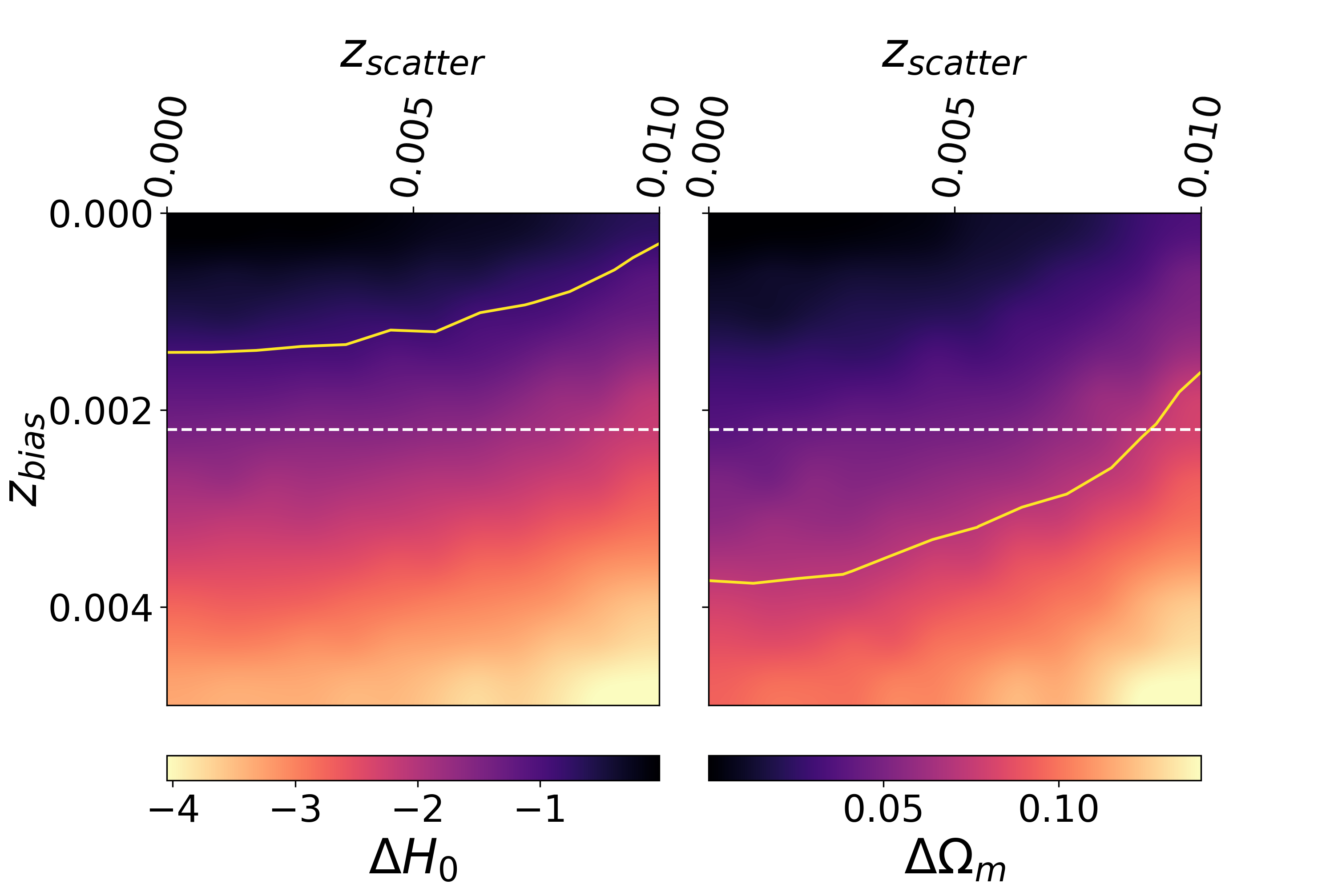}
    \caption{Estimated bias induced in $H_0$ and $\Omega_m$ for SNe with not-hostz redshift distribution for simulated trials using different statistical uncertainty and bias in redshift. Increased statistical uncertainty and bias will produce more biased cosmological parameters. The orange contours in each panel indicate the offsets in $H_0$ and $\Omega_m$ between the hostz and not-hostz samples (Table \ref{table:bestfit}) The white, dashed line is drawn at the SNz bias reported in \citet{Sako2018}.  Unlike this simulation, in practice individual objects will have different statistical uncertainties and possibly different biases.}
    \label{fig:scatter-bias}
\end{figure} 

\subsection{Differences Between Supernovae with High-Quality and Low-Quality Redshifts} 
\label{subsec:samplediffs} 

A significant concern with subdividing the Pantheon sample is that it implicitly assumes there are no relevant hidden variables which differ across the two samples.  The assumption that a Type Ia supernova is a standard candle implies that no such hidden variable should exist.  However, there is certainly a difference in the host galaxies of, e.g., the hostz and not-hostz samples.  For example, it is easier to resolve the host and supernova independently for galaxies with more extended morphology, such as spirals.  If there were a correlation between intrinsic Type Ia supernova luminosity and some of these properties, it would also produce a discrepancy between the high-quality and low-quality samples.  Of course, if the tension between hostz and not-hostz samples arises from such a correlation, it would also potentially invalidate the use of supernovae as standard candles.  However, considerable effort \citep{March} has gone into including these effects in the luminosity distance error budget.  Hosts for which a host redshift can be determined also often have host galaxy noise contributing to supernova photometry, particularly towards high redshift where the host and supernova are not resolved independently.

Perhaps the most significant difference between the two samples lies in the redshift distribution (Fig. \ref{fig:residualcomp}, bottom).  Lower-redshift hosts are far more likely to have been resolved independently or to have had existing host spectra.  A true redshift dependence in Type Ia luminosities would similarly invalidate supernova cosmology, but has also been extensively studied (cf. \citealt{Sapone2020}) and is an unlikely explanation. However, one indicator of a non-$\Lambda$CDM cosmology would also be that one cannot fit the same $\Lambda$CDM parameters across all redshifts. 

In order to test this, samples of the same size and redshift distribution as the not-hostz sample were drawn from the hostz sample with replacement.  The resulting best-fit resampled parameters averaged $\Omega_m = 0.293$, with a standard deviation of $0.017$. The distribution is consistent with the hostz central value of $\Omega_m = 0.270$ but not the not-hostz central value of $\Omega_m = 0.374$.  Because there are very few objects at $z < 0.1$ in the not-hostz sample, the resampled central value is somewhat sensitive to the exact distribution; omitting the ambiguous objects\footnote{As in Sec. \ref{subsec:sampledefinition}, in a small number of cases authors describe spectra as having a mixture of SN and host lines, or being SN-dominated with one or two host features.  For this analysis, such objects are part of neither the hostz nor SNz samples.} and resampling the hostz catalog using only objects flagged as SNz instead produces a mean $\Omega_m = 0.276$ and standard deviation of $0.020$, which is also consistent with the hostz value of $\Omega_m$ but not the not-hostz measurement.  Thus, the difference between high-quality and low-quality samples is not created by the redshift distribution.  

An additional possibility would be that the best-fit $\Lambda$CDM cosmological parameters truly differ between samples with different redshift distributions.  That is, $\Lambda$CDM might be the wrong cosmological model, in which case the same $\Lambda$CDM parameters would not be able to fit the luminosity distance-redshift relation at all redshifts.

However, the same $\Lambda$CDM cosmology is consistent with hostz-based supernova samples at both high and low redshift.  Similarly, a different $\Lambda$CDM cosmology is consistent with SNz-based supernova samples are both high and low redshift.  The difference between these cosmologies may stem from systematic errors in SNz measurements, or perhaps from a systematic difference in host galaxy properties, but it is not merely a result of the redshift distribution.

An additional significant difference between the two subsamples may come from Malmquist bias \citep{Eddington1913,Malmquist1922,Malmquist1925}.  Towards higher redshift, when resolving the supernova and host independently becomes more difficult, the brightest supernovae are more likely to dominate the emission than fainter ones, resulting in a SNz rather than a hostz.  They are also likely to be bluer than supernovae with hostz. Similarly, it is harder to measure precise redshifts for faint galaxies, which also be bluer and less massive. A change in sample composition requires recalibrating for sample completeness to properly account for these effects.  This is a key reason that the subsample fits in this work cannot be used as precision cosmological tests, but rather are useful diagnostics of which effects are most significant.

\section{Discussion}
\label{sec:discussion}

In this work, we have evaluated the effects of several potential issues with redshift measurements and uncertainties in Type Ia supernova samples.  Neglecting low redshift uncertainties $\gtrsim 10^{-3}$ or systematic errors $\gtrsim 10^{-3}$ at any redshift leads to biased cosmological parameters.  This is of particular concern because the Pantheon catalog is compiled from redshifts measured using two very different techniques: just under 70\% of the catalog uses redshifts determined from a spectrum of the host galaxy (hostz), and the remainder instead relies on fits to a supernova-dominated spectrum (SNz).  A separation into hostz and SNz subsamples yields statistically significantly different best-fit cosmological parameters.  

It is tempting to only use the higher-precision hostz subsample for cosmology.  However, there are also systematic differences in the host galaxy properties of the two subsamples \citep{Kelly, Lampeitl, Sullivan}.  If these differences were to affect the luminosity calibration of Type Ia supernovae, it could also produce tension between hostz and SNz best-fit parameters like the one found here.  In that case, using hostz alone would not fix the problem.  Further, the Pantheon calibration relies on the entire sample, invalidating the use of any subsample for precision cosmological measurements without first recalibrating only that subsample with new bias estimates.

Nevertheless, it is clear that the most significant effect of moving to a hostz-only sample will be to decrease measurements of $\Omega_m$ and the matter density $\Omega_m h^2$, where $h = H_0 / 100\textrm{ km/s/Mpc}$.  An intriguing possible outcome would be the one suggested by the best-fit hostz and Known-Error parameters for flat $\Lambda$CDM in Table \ref{table:bestfit} (although we again emphasize that subsamples are not properly calibrated for precision cosmological measurements.)  In our pilot study, there is a tension between the best-fit values for $\Omega_m$ and $H_0$ when compared to the best-fit values from CMB measurements.  However, both subsamples produce $\Omega_m h^2$ consistent with the Planck measurement of $0.142 \pm 0.001$.

$\Omega_m h^2$ is one of the best-measured quantities derived from the CMB power spectrum since it is directly related to the sound horizon as determined from the location of the first acoustic peak.  In the lower-redshift measurements discussed in this paper, $H_0$ and $\Omega_m$ are determined independently and have different sources of systematic error.  For example, changing the luminosity calibration from the cosmological distance ladder will alter the value obtained for $H_0$ but not for $\Omega_m$.  Thus, if it were to turn out that local and CMB values of $\Omega_m h^2$ agree while measurements of $H_0$ and $\Omega_m$ are different, it would strongly constrain possible theoretical explanations.  A comparison of local $\Omega_m h^2$ measurements with Planck therefore presents an intriguing diagnostic for future measurements. 

In particular, the possibility that there might be tension in $H_0$ and $\Omega_m$ but not $\Omega_m h^2$ lends itself to a different class of theoretical models than tension in $H_0$ alone.  To change $H_0$ at late times, a natural model would have been to change the dark energy equation of state, something which would only take effect at low redshift, as dark energy becomes significant.  Such a model would not change $\Omega_m$, so now a second effect would need to be introduced.  At early times, however, the two measurements are instead $\Omega_m h^2$ and a structure formation-based measurement (BAO, additional CMB acoustic peaks, etc.) to separate $H_0$ and $\Omega_m$.  Thus, a model could explain tension in both parameters with the introduction of just one change to dark matter which alters early-Universe structure formation.  This is particularly intriguing in light of observations of massive, high-redshift galaxies which hint at additional tension with $\Lambda$CDM \citep{Steinhardt2016,Behroozi2018}.

The natural next step is increase the size and completeness of the hostz samples.  Adding hundreds of new objects does not require waiting to discover hundreds of additional supernovae.  Rather, it is only necessary to remeasure redshifts for objects without host redshifts, which can be done immediately because the supernovae will already have faded.  In other cases, the data exist but are merely insufficiently documented, and a re-analysis of existing spectroscopy will suffice.  

If a hostz can be measured for the entire Pantheon sample, then it will be possible to include the entire catalog, at all redshifts, with identical completeness to the current sample.  Some recalibration will still be required even with new redshifts.  For example, changes to the redshift of an object also change the K correction used in determining the distance modulus.  However, measuring host redshifts for the remaining 346 objects in the Pantheon sample is a step which can be taken immediately, and comparing those redshifts with existing measurements is an excellent way to determine whether SNz (and the resulting cosmological parameters) are flawed. 

This will allow us to conduct an additional necessary test which cannot be performed from current catalogs.  The SNz sample exists not by choice, but out of necessity because the spectral observations used to confirm a Type Ia supernova could not resolve the host independently.  This means Pantheon supernovae using SNz also should lie in galaxies with systematically different morphology, radius, etc. than the hostz sample.  If the luminosity of a Type Ia supernova were correlated with these properties, it could produce a similar effect to a systematic error in SNz redshifts.  If so, using hostz for these galaxies will not change cosmological fits, so this can also be tested by once the missing host redshifts are measured.

This issue is particularly critical because current and near-future surveys will need to measure redshift rapidly in order to produce large samples.  Thus, the effort needed to produce host redshifts from followup spectroscopy will grow rapidly and it will tempting to rely increasingly on SNz towards high redshift.  If new SNz behave similarly to those in the Pantheon sample, the resulting measurement of $\Omega_m$ should increase as the fraction of SNz goes up.  For large galaxy catalogs, a similar problem resulted in a shift to photometric redshifts, which are even less precise and more prone to systematic errors than SNz \citep{Bezanson2016}.  Photometric redshifts might create a larger bias, but without further study the direction of that bias in $\Omega_m$ is not known.  

In summary, at present supernova cosmology is effectively not one technique for measuring the composition of the Universe, but two closely related techniques.  Those techniques appear to yield conflicting cosmological parameters for current models.  Thus, combining the two into a mixed sample as has been done to this point is very likely flawed.  Unfortunately, the superior technique requires spectra of host galaxies and is more expensive observationally.  However, in an era of precision supernova cosmology, this additional expense appears necessary in order to produce meaningful results, both to complete current catalogs and for future surveys.

The authors would like to thank Nikki Arendse, Tamara Davis, Johan Fynbo, Jitze Hoogeveen, Rasmus Nielsen, Saul Perlmutter, Steve Rodney, David Rubin, David Spergel, Paul Steinhardt, Sune Toft, Darach Watson, and Radek Wojtak for useful discussions.  CLS is supported by ERC grant 648179 "ConTExt".  BS is supported by the MIT International Science and Technology Initiatives (MISTI) Denmark program.  The Cosmic Dawn Center (DAWN) is funded by the Danish National Research Foundation under grant No. 140. 

\bibliographystyle{mnras}
\bibliography{refs.bib} 

\appendix

\section{Details of Corrections to Pantheon CMB-centric Redshifts and sample Selection}
\label{app:zcmb}

\subsection{Conversion Between Heliocentric and CMB Frames}

The Pantheon conversion between CMB-frame redshift \zc~and heliocentric \zh~has been performed using low-redshift approximations (D. Scolnic, personal communication; \citealt{Davis2019}).  Specifically, the Pantheon combinations of \zc~and \zh~can be reproduced under the assumption that $z = v/c$, where $v$ is the recessional velocity and $c$ is the speed of light.  This approximation is only valid at very small z, but appears to be used even for $z > 1$.

A more proper approach is to note that in general \citep{Davis2004}, 
\begin{equation}
(1 + z) = (1 + z_{\textrm{hel}})(1 + z_{dipole,\perp}) = 1 + z_{\textrm{hel}} + z_{dipole,\perp} + z_{\textrm{hel}}z_{dipole,\perp}.
\end{equation}

The Pantheon conversion omits this last term, which can become significant (e.g., 20\% the size of the dipole itself at $z=0.2$).

At low redshift, a similar approach was used for including a bulk flow correction from the 2M++ catalog \citep{Carrick2015}. A full relativistic treatment of velocities and redshift reproduces the \zh~to \zc~conversion in the JLA catalog.  However, the Pantheon \zc~and \zh~are inconsistent with this treatment and therefore are adjusted in the catalog presented here.

We reconvert between \zh~and \zc~as required, then add the appropriate 2M++ bulk flow correction.  The Pantheon catalog has been assembled from a variety of sources.  In some cases the approximate conversion is from a measured \zh~to \zc, and the adjustment alters \zc~and therefore the resulting fit.  In other cases, the approximate is from the \zc~in a different catalog back to \zh, and therefore \zc~and the resulting fit will not be affected.  

\subsection{SNLS}

The Pantheon heliocentric redshifts approximately match the original redshifts measured by SNLS, with the exception of artificial rounding in the sixth decimal place.  The catalog used here takes \zh~directly from \citet{Guy2010}, then converts from \zh~to \zc.  The result is a \zh~very similar to the Pantheon \zh, but a \zc~that differs more significantly from the Pantheon \zc~for each object.  

The original SNLS paper has both labels for the type of redshift as well as the errors. Therefore, all of these objects are in the hostz sample. Although hostz redshift uncertainties are reported individually for each object, SNz uncertainties are not. Therefore, only supernovae with host redshifts are included in the known-error sample.

\subsection{SDSS}
SDSS appears to report heliocentric redshifts, which match the redshifts given in the JLA data set for most objects in both catalogues.  JLA uses \zh~from SDSS \citet{Smith2012}, including a table\footnote{https://classic.sdss.org/supernova/snlist\_confirmed.html} with some updated redshifts.  JLA then converts \zh~to \zc~correctly.  Pantheon uses the value of \zc~from JLA, then converts back to \zh.  Thus the Pantheon \zc~values will be generally correct, but their errors in redshift conversion mean that the Pantheon \zh~will systematically disagree with both JLA and SDSS.  In addition, there are some host galaxies for which redshifts were remeasured in later SDSS observations, and those redshifts were included in the Pantheon but not JLA samples.  This also appears to be the root cause of the largest discrepancies in \zh~between JLA and Pantheon reported in \citet{Rameez2019}.

The catalog used here uses \citet{Smith2012}, \citet{Sako2018}, and SDSS SkyServer DR16 to draw \zh, then converts to \zc.  The \zc~will be similar to Pantheon in cases where \zc~was drawn from JLA, but the \zh~will systematically disagree. For these objects, the distance modulus and their uncertainties from the original Pantheon catalog were used to find the best-fit $H_0$. The types of redshifts and uncertainties are also taken from \citet{Smith2012}, \citet{Sako2018}, and SDSS SkyServer DR16, and 307 of these objects are included in the hostz sample. The host galaxy redshifts are also included in the known-error sample.

\subsection{CSP}
Supernovae from the Carnegie Supernova Project \citep{Contreras2010} have \zh~reported as having been calculated from NED for host galaxies. CSP \zh~matches the Pantheon catalog \zh~to within rounding.  The catalog presented here uses the Pantheon \zh~as canonical, then converts to \zc.  These values are reported in the catalog, along with the NED uncertainties. These objects are in the hostz and known-error samples.  

The redshift uncertainty for 2MASX J09151727-2536001SN, host of 2006lu, could not be found in NED, and therefore 2006lu is not included in the known-error sample.

\subsection{CfA1-4} 
Where objects appear in both the Pantheon and JLA catalogs, the \zh~agree but \zc~systematically disagree.  The catalog presented here uses the Pantheon \zh~as canonical, then converts to \zc.  The \zh~will therefore be identical to Pantheon, but the \zc~will systematically disagree. Although uncertainties are not reported for these objects, these redshifts can be found as host galaxy redshifts on NED. Therefore, these objects are part of the hostz sample, where the uncertainties in the second sample are set to $10^{-4}$. Some of the CfA3 objects were located in SDSS SkyServer DR16. For these objects, the SDSS reported redshifts and uncertainties are presented, and these objects are in both the hostz and known-error samples. 1995ak from CfA1 is only in the hostz sample.

\subsection{Pan-STARRS}

The original Pan-STARRS spectra have not been made available, but it is assumed that the \zh~reported in the Pantheon datatables are canonical.  These redshifts are then converted to the \zc~in our catalog.  The \zh~will therefore be identical to Pantheon, but the \zc~will systematically disagree. These objects are in the hostz sample but not known-error, as exact uncertainties are not available. The objects that have spectra with host galaxy emission lines in an SN spectrum are not in any of the samples.

\subsection{Hubble Programs (CANDELS, CLASH, GOODS, SCP)}

The redshifts given by HST are \zh, listed individually with their sources in \citet{Rodney2014} and \citet{Graur2014}.  However, the same values are given in the Pantheon catalog as \zc, not \zh.  Pantheon then incorrectly converts \zc~to \zh.  The corrected \zh~used here will therefore be identical to the \zc~given in Pantheon for each supernova, but different from the \zh~reported by Pantheon.  The \zc~will also be different, because it is then converted from the correct \zh. 

Due to the large uncertainties reported in the CLASH/CANDELS paper for all of the objects except Primo, objects from these surveys are only part of the hostz sample. Primo is in the known-error sample. 

Although redshift errors are not reported for the GOODS or SCP objects, the spectra for these objects are shown in the survey paper. Therefore, these objects are included in the hostz sample, but not the known-error one. Additionally, there are five objects from the GOODS survey where the redshifts have been determined from both broad supernova features and narrow host emission lines. These objects are not in any of the samples.

\subsection{Updated Pantheon Catalog}

After including all of the effects and labeling subsamples as described in this work, an updated Pantheon catalog is produced.  All objects are labeled with redshift uncertainties, but only the ones flagged as part of the known-error sample are individually reported.  The remainder use estimates from the authors of individual source catalogs.

\begin{table}[ht]
    \centering
    \begin{tabular}{l l l l l l l l l l l}
        \hline
         Name & Mu & Dmu & \zh & \zc & Error & Ra & Dec & Survey & Type & Known-error \\
         \hline
         03D1au & 42.2688 & 0.1184 & 0.504300 & 0.503089 & 0.0005 & 36.04320908 & -4.03746891 & SNLS & H & yes \\
         03D1ax & 42.2277 & 0.1173 & 0.496000 & 0.494801 & 0.001 & 36.09728622 & -4.720774174 & SNLS & H & yes \\
         03D1co & 43.3714 & 0.1970 & 0.679000 & 0.677668 & 0.001 & 36.56774902 & -4.935050011 & SNLS & H & yes \\
         ... & ... & ... & ... & ... & ... & ... & ... & ... & ... & ... \\
         \hline
    \end{tabular}
    \caption{Updated Pantheon catalog with the corrections described in Appendix \ref{app:zcmb} and the sample distinctions described in \S~\ref{sec:hostzsnz}. }
    \label{tab:catalog}
\end{table}

\label{lastpage}
\end{document}